\documentclass[aps, prd, nofootinbib,superscriptaddress,preprintnumbers,letterpaper,natbib,nofootinbib,twocolumn]{revtex4-1}
\pdfoutput=1
\usepackage{amssymb,amsmath,latexsym,mathrsfs}
\usepackage{url}
\usepackage{enumitem}
\usepackage{graphicx}
\usepackage[usenames,dvipsnames]{xcolor}
\usepackage[breaklinks,colorlinks,urlcolor=blue,citecolor=blue,linkcolor=magenta]{hyperref}
\usepackage{multirow}
\usepackage{rotating}
\usepackage[caption=false]{subfig}
\usepackage{scalerel}
\usepackage[dvipsnames]{xcolor}
\usepackage{cleveref}

\begin{document}
\title{Neutron Lifetime Anomaly and Big Bang Nucleosynthesis}
\author{Tammi Chowdhury}
\email{Tammi.Chowdhury@carleton.ca}
\affiliation{Carleton University  1125 Colonel By Drive, Ottawa, Ontario K1S 5B6, Canada}

\author{Seyda Ipek}
\email{Seyda.Ipek@carleton.ca}
\affiliation{Carleton University  1125 Colonel By Drive, Ottawa, Ontario K1S 5B6, Canada}

\begin{abstract}
We calculate the Big Bang Nucleosynthesis abundances for helium-4 and deuterium for a range of neutron lifetimes, $\tau_n = 840 - 1050$~s, using the state-of-the-art Python package \textsc{PRyMordial}. We show the results for two different nuclear reaction rates, calculated by NACRE II~\cite{Xu:2013fha} and the PRIMAT~\cite{Pitrou:2020etk} collaborations. 

\end{abstract}

\maketitle
%\tableofcontents

%%%%%%%%%%%
\section{Introduction}
%%%%%%%%%%%
The primordial abundances of light elements are the earliest measurements of the history of our universe. During the Big Bang Nucleosynthesis (BBN) era, corresponding to temperatures $T\sim O({\rm MeV})$, light elements such as helium, deuterium, tritium, lithium and beryllium were produced as a result of nuclear reactions predicted in the Standard Model (SM). The final abundances depend on parameters such as the number of relativistic species ($N_{\rm eff}$), baryon-to-photon ratio $\eta$, neutron-proton mass difference and the neutron lifetime $\tau_n$. These parameters can all be calculated within the SM or measured in SM processes, except $\eta$, whose value can be determined from  BBN and the cosmic microwave background (CMB) individually. Given the SM predictions for BBN abundances, the primordial abundance measurements can then be used to constrain new physics beyond the SM. 

There has been a consistent discrepancy between two different methods of measuring the neutron lifetime in the last decades, suggesting possible new physics. As will be described in the next section, ``bottle" experiments give $\tau_n=878.4\pm 0.5$~s~\cite{Workman:2022ynf} while ``beam" experiments find a higher value of $\tau_n=888.45\pm 1.65$~s~\cite{Rajan:2018hii}. Neutron lifetime is an important ingredient in calculating the SM predictions of BBN abundances. Thus, BBN measurements can be compared to theoretical expectations to extract the neutron lifetime. This possibility has been studied in the literature~\cite{Mathews:2004kc, Salvati:2015wxa}. We revisit this idea in light of newest neutron lifetime measurements, BBN observations as well as updates on nuclear rates that go into BBN calculations. We use a recent Python package \textsc{PRyMordial}~\cite{Burns:2023sgx}. (See \cite{Burns:2022hkq} for details about the program.) Our results are shown in \Cref{fig:abundances}.

%%%%%%%%
\section{Neutron Lifetime Anomaly}
%%%%%%%%
In the SM the neutron decays to a proton, an electron and antineutrino via electroweak interactions, $n\to p\, e^-\,\bar{\nu}$\footnote{A small fraction, about $10^{-3}$, of decays produce final states with a photon. Even more rarely there will be decays in which electron binds to the proton to form a hydrogen atom.}. The SM decay rate is~\cite{PhysRevLett.96.032002}
\begin{align}
    \tau_n = \frac{4908.7(1.9)~{\rm s}}{|V_{ud}|^2(1+3g_A^2)}\,,
\end{align}
where $V_{ud}=0.97370 \pm 0.00014$ is an element of the Cabibbo-Kobayashi-Maskawa (CKM) quark mixing matrix and $g_A \equiv G_A/G_V \simeq 1.27$ is the ratio of axial  and vector couplings. 

Neutron lifetime is measured using two experimental methods. In the \emph{bottle method} ultracold neutrons (UCNs) are trapped in a container. After a time comparable to the expected neutron lifetime, the remaining neutrons are counted. In the \emph{beam method} a slow neutron beam decays in an experimental volume. The decay products, protons and electrons, and the remaining neutrons are captured. Given the neutron flux, the beam method measures the amount of neutrons that decayed into protons in a decay volume. This then can be translated to a neutron lifetime, assuming neutrons only decay in this way.  (See \cite{atoms6040070} for a historical review of neutron lifetime measurements.) These two methods give
\begin{align}
    \tau_n^{\rm bottle} = 878.4\pm 0.5~{\rm s}\,~~~~ \tau_n^{\rm beam} = 888.45\pm 1.65~{\rm s}\,,
\end{align}
with more than $4\sigma$ discrepancy between the two measurement methods. This discrepancy could be due to unaccounted-for systematic errors in one or both methods or it could be due to new physics beyond the SM~\cite{Fornal:2018eol,PhysRevLett.124.219901,PhysRevD.103.035014,TAN2019134921,Cline:2018ami}. For example, if neutron decays to dark matter, beam experiments, which count the decay products, would interpret that as a longer (SM) neutron lifetime while bottle experiments, which are blind to different decay channels, will still measure the total neutron lifetime.

Very recently space-based measurements of the neutron lifetime emerged~\cite{PhysRevResearch.2.023316,PhysRevC.104.045501}. Neutrons are freed from the surface of planetary objects by galactic cosmic rays. These neutrons thermalize with the atmosphere with velocities of a few km/s. Spacecrafts that are $O(1000~{\rm km})$ above the surface of the planetary object can be sensitive to the neutron lifetime by measuring the neutron flux from the object's surface, given its elemental composition. NASA's MESSENGER spacecraft performed such a measurement during flybys of Venus and Mercury~\cite{PhysRevResearch.2.023316} and recently the Lunar Prospector Mission measured the neutron lifetime ~\cite{PhysRevC.104.045501}:
\begin{align}
    \tau_n^{\rm space}=887\pm 14~{\rm s}
\end{align}
These error bars are an order of magnitude larger than the other two methods. Still, we include this measurement in our results as a very exciting opportunity for a new kind of neutron lifetime measurement.

%%%%%%%%%%%
\section{BBN Abundances}
%%%%%%%%%%%
\begin{figure}[t]
    \centering
    \includegraphics[width=\columnwidth]{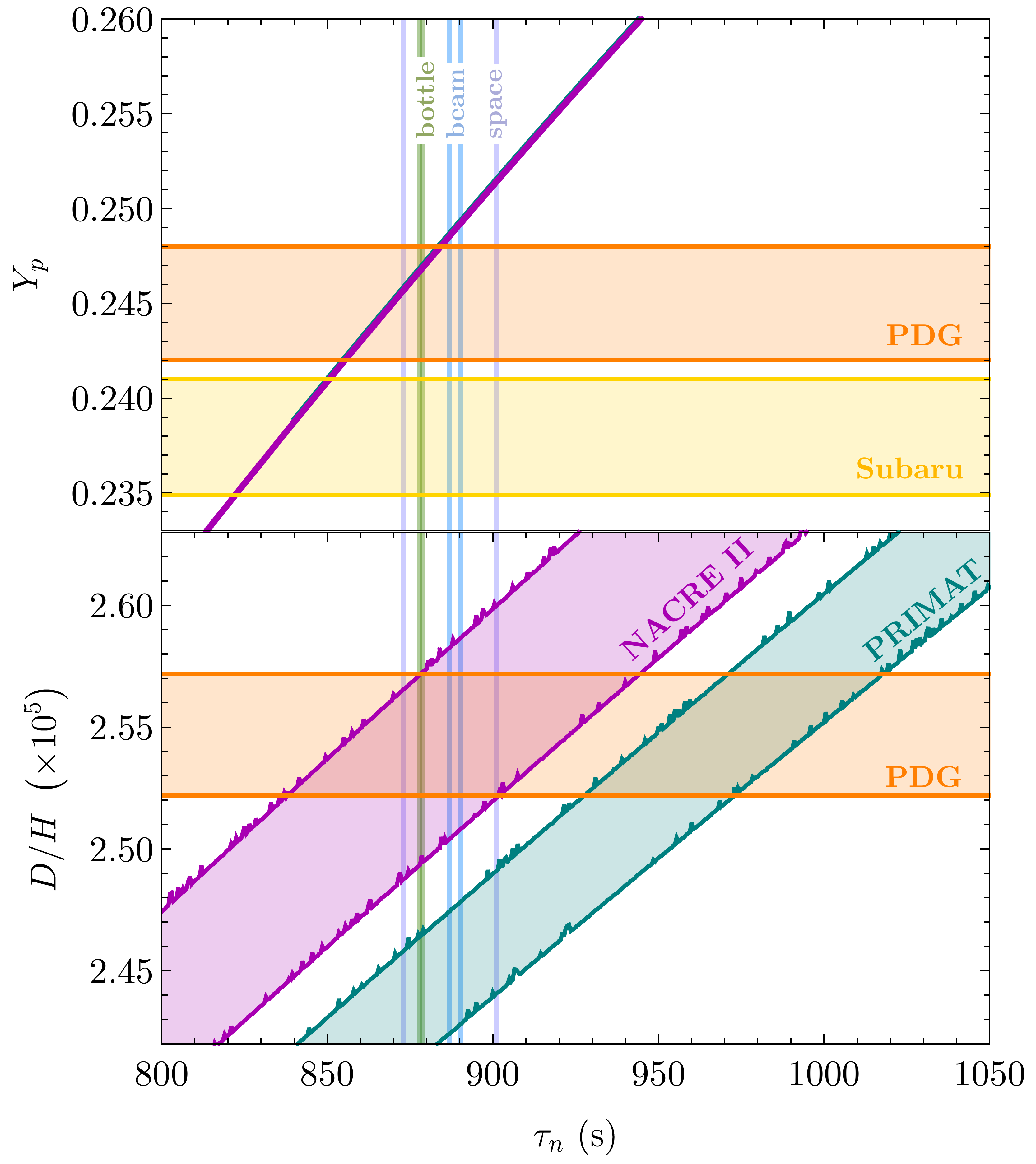}
    \caption{Helium-4 mass fraction ($Y_p$) and relative deuterium abundance (D/H) versus neutron lifetime, $\tau_n$. (The little spikes on the deuterium lines are due to numerics and are not physical.) The shaded regions for the theoretical calculations correspond the $1\sigma$ error bars on the CMB measuremnt of baryon-to-photn ratio, $\eta=(6.105\pm0.057)\times 10^{-10}$. ($D/H$ is lower for higher $\eta$.) The two different abundance predictions, NACRE II and PRIMAT, are explained in the text. Nuclear reaction uncertainties, which are expected to be at $1\%$ level for $D/H$, are not included.  Horizontal, orange shaded regions are observed primordial abundances given by PDG~\cite{Workman:2022ynf}. For helium-4, we also show a recent measurement by the Subaru telescope~\cite{Matsumoto:2022tlr} in yellow. The vertical lines are averages of the neutron lifetime determined by beam, bottle, and space experiments.}
    \label{fig:abundances}
\end{figure}

The earliest observations we have about our universe are from the era of BBN, when the temperature was $O(1~{\rm MeV})$. The abundances of primordial light elements can be very sensitive to disturbances to SM processes during BBN.
Consequently, comparison of primordial element observations to theoretical expectations is a good probe of physics beyond the SM in the early universe.

BBN starts when the weak interactions keeping neutrons in equilibrium, \emph{e.g.} $p+e^-\longleftrightarrow n +\nu_e$, freeze out. Since this is the same interaction that facilitates neutron decay, the rate for this process is related to the neutron lifetime. In order to find the freeze-out temperature, one compares this scattering rate, $\sim G_F^2 T^5$, to the Hubble rate, $H(T) = 1.66\sqrt{g_\ast}\,T^2/M_{\rm Pl}$, where $G_F \simeq 1.16\times 10^{-5}~{\rm GeV}^{-2}$ is the Fermi constant, $M_{\rm Pl}=1.2\times 10^{19}~$GeV is the Planck mass and $g_\ast$ is the relativistic degrees of freedom at a temperature $T$. In the SM, at the neutron freeze-out temperature $T_n\sim 0.75~$MeV, only contributions to $g_\ast$ are from photons, electrons, positrons and neutrinos. There are many hints that the effective number of relativistic neutrino species, $N_{\rm eff}$, might be different than the SM expectation during BBN~\cite{Burns:2022hkq, Matsumoto:2022tlr}. Here we use the SM value: $N_{\rm eff} = 3.044$. 

At $T_n$, the ratio of number densities of neutrons and protons is fixed at $n_n/n_p|_{T=T_n}= \exp\left(-\Delta m/T_n\right)$, where $\Delta m= m_n-m_p \simeq 1.3~$MeV is the mass difference between the neutron and the proton. After the neutron freeze-out, proton number does not change while neutrons can decay with a lifetime $\tau_n$. As the temperature further cools, protons and the remaining neutrons bind into deuterium via the reaction $p+n\to D+\gamma$. Nucleosynthesis starts with the formation of deuterium at $T_{\rm NS}\sim 100~$keV~\footnote{Even though $T_n$ is smaller than the deuteron binding energy, $\sim 2.2$~MeV, the high energy tail of the photon distribution still can disassociate deuteron effectively. Nucleosynthesis is delayed until this process falls out of equilibrium. This is called the \emph{deuterium bottleneck}.}, setting off a series of nuclear reactions that eventually produces heavier elements up to lithium-7 and beryllium-7.    

After being produced, deuterium is burned into tritium and helium-3 through the reactions
\begin{align}
    \begin{split}
        D+p \to {}^3{\rm He} +\gamma\, ,& ~~~
        D + D \to {}^3{\rm He} + n\,, \\
        D + D &\to T + p\,.
    \end{split}\label{eq:Dburning}
\end{align}
Deuterium burning stops at $T_D \sim 70$~keV, leaving a small deuterium abundance which is traditionally described as a relative abundance, $D/H$.  Deuterium is only destroyed in stars, and hence its observation in the universe is purely primordial. The PDG average~\cite{Workman:2022ynf} of the most recent measurements~\cite{Pettini_2001, DOdorico:2001vgo,2010MNRAS.405.1888S, 2011Sci...334.1245F, 2012A&A...542L..33N, 2018MNRAS.477.5536Z, Cooke:2013cba, Balashev:2015hoe, Cooke:2016rky, Cooke:2017cwo, Riemer-Sorensen:2014aoa, Riemer-Sorensen:2017pey} give
\begin{align}
    D/H^{\rm PDG}= (25.47 \pm 0.25)\times 10^{-6}
\end{align}
This relative abundance decreases as the baryon-to-photon ratio increases\footnote{The dependence can be numerically obtained as $\eta^{-1.6}$~\cite{Yeh:2020mgl}.} $\eta$ and is a good candidate to confirm the CMB measurement of this quantity:
\begin{align}
    \eta_{\rm CMB}=(6.105\pm0.057)\times 10^{-10}
\end{align}
We use the above value of $\eta$ in \Cref{fig:abundances}. As seen in \Cref{fig:abundances}, $D/H$ also has a mild dependence on the neutron lifetime.

Theoretical predictions of the deuterium abundance rely on experimental measurements of the rates of the reactions given in \Cref{eq:Dburning} at energies relevant for BBN ($E\simeq 40~{\rm keV}-400~$keV). The latest and most accurate measurement of $D+p\to {}^3{\rm He} +\gamma$ comes from the LUNA collaboration~\cite{Mossa:2020gjc}. (See \cite{Tisma:2019acf} for a preceding measurement.) Currently, theoretical error in $D/H$ prediction is driven by the other two reactions. Given the existing, albeit limited, experimental data, there are different approaches to data selection and analysis to deduce the nuclear rates that go into abundance calculations. One approach is to find a polynomial fit for the energy dependence using a $\chi^2$-analysis. (See, e.g. \cite{Yeh:2020mgl,Pisanti:2020efz,Mossa:2020gjc} for recent such analyses which agree with each other.) Another approach is to use theoretical \emph{ab initio} calculations as a starting point to extract the relevant rates from experimental data. NACRE II~\cite{Xu:2013fha} and PRIMAT~\cite{Pitrou:2018cgg,Pitrou:2020etk} follow this approach, but differ in which datasets to include in their analyses. When the new LUNA results are included the $D/H$ got smaller but \cite{Yeh:2020mgl,Pisanti:2020efz,Mossa:2020gjc} still agree with NACRE II calculatios at about $1\%$ level. PRIMAT result using the same LUNA rate seems to be an outlier. This discrepancy suggests that the difference stems from the other two rates. (See \cite{Pitrou:2021vqr} for a short discussion.)

The most stable primordial nuclei is helium-4 and it is produced via the burning of helium-3 and tritium:
\begin{align}
    {^3}{\rm He} + D \to {^4}{\rm He} +p\,, ~~~ T+D \to {^4}{\rm He} +p\,.
\end{align}
Practically all neutrons that make it to the start of nucleosynthesis end up in helium-4, resulting in the mass fraction 
\begin{align}
    Y_p&= \frac{m_{{^4{\rm He}}}\,n_{{^4{\rm He}}}}{m_p n_p+m_n n_n} \simeq \left.\frac{2}{1+n_p/n_n}\right|_{T=T_{\rm NS}} \notag \\
    &\simeq \frac{2}{1+e^{\Delta m/T_n}e^{t_{\rm NS}/\tau_n}} \simeq 25\%\,,
\end{align}
 where $t_{\rm NS}\simeq 200~$s is the time at $T=T_{\rm NS}$. \Cref{fig:abundances} shows this strong dependence on $\tau_n$. 
 
 Measuring the primordial abundance of helium-4 is more complicated than deuterium since helium-4 is abundantly produced in stars. In order to determine the primordial abundance, metal-poor galaxies, where star formation is low, are chosen. The PDG average~
 \cite{Workman:2022ynf} of the current measurements~\cite{Aver:2020fon, Valerdi:2019beb, Fernandez:2019hds, Kurichin:2021ppm, 2020ApJ...896...77H, 2021MNRAS.505.3624V} is 
 \begin{align}
     Y_p^{\rm PDG}=0.245\pm 0.003~.
 \end{align}
Not included in the above average is a very recent result from the Subaru Telescope~\cite{Matsumoto:2022tlr}, which is obtained from 10 extra-metal-poor galaxies. Combined with earlier data from another 3 galaxies, this new result is lower than previous ones: 
\begin{align}
    Y_p^{\rm Subaru} = 0.2379^{+0.0031}_{-0.0030}\,.
\end{align}
As can be seen in \Cref{fig:abundances}, this result disagrees with the PDG average at a little more than $1\sigma$ level. 

There is also leftover primordial helium-3, with a relative abundance comparable to $D/H$. The claimed measurements of primordial helium-3 have been disputed. Lithium-7 abundance is predicted to be $O(10^{-9})$ and possibly anomalously large compared to observations for the expected baryon-to-photon ratio. We do not include these abundances here.

%%%%%%%%%%%%%%
\section{Discussion and Outlook}
%%%%%%%%%%%%%%
Our results are summarized in \Cref{fig:abundances}. While the beam result for the neutron lifetime is slightly disfavored by the helium-4 abundance, current observations  and predictions obtained from  \textsc{PRyMordial} agree with neutron lifetime results from bottle experiments. However, the most recent primordial helium-4 measurement claim by the Subaru Telescope would require a much lower neutron lifetime for the $N_{\rm eff}$ and $\eta$ values we use. 

When NACRE II results are used for nuclear reaction rates, deuterium abundance agrees with both beam and bottle results with a preference for lower $\eta$ values. PRIMAT-driven deuterium abundance predictions are in tension with neutron lifetime measurements for $N_{\rm eff}=3.044$. 

The era of BBN has been very important for driving our understanding of the SM in the early universe as well as guiding us scrutinize new physics that can alter the primordial production of light elements. Here we emphasize the importance of continued BBN studies for the long-lasting question of the neutron lifetime.

\section*{Acknowledgements}
We thank Anne-Katherine Burns, Tim Tait and Mauro Valli for sharing an early version of \textsc{PRyMordial} and Mauro Valli for his help in using the program. This work is supported in part by the Natural Sciences and Engineering Research Council (NSERC) of Canada.

Competing interests: The authors declare there are no competing interests.

Data generated or analyzed during this study are provided in full within the published article.

\bibliography{refs}
\end{document}